# Evaluation of WRF-Sfire Performance with Field Observations from the FireFlux experiment

Adam K. Kochanski[1], Mary Ann Jenkins[1,4], Jan Mandel[2], Jonathan D. Beezley[2], Craig B. Clements[3], Steven Krueger[1]

[1]Department of Atmospheric Science, University of Utah, Salt Lake City, UT
[2]Department of Mathematical and Statistical Sciences, University of Colorado Denver, Denver, CO, USA
[3]Department of Meteorology and Climate Science, San José State University, San José, CA
[4]Department of Earth and Space Science, York University, Toronto, ON, Canada

**Abstract**

This study uses in-situ measurements collected during the FireFlux field experiment to evaluate and improve the performance of coupled atmosphere-fire model WRF-Sfire. The simulation by WRF-Sfire of the experimental burn shows that WRF-Sfire is capable of providing realistic head fire rate-of-spread and the vertical temperature structure of the fire plume, and, up to 10 m above ground level, fire-induced surface flow and vertical velocities within the plume. The model captured the changes in wind speed and direction before, during, and after fire front passage, along with arrival times of wind speed, temperature, and updraft maximae, at the two instrumented flux towers used in FireFlux. The model overestimated vertical velocities and underestimated horizontal wind speeds measured at tower heights above the 10 m, and it is hypothesized that the limited model resolution over estimated the fire front depth, leading to too high a heat release and, subsequently, too strong an updraft. However, on the whole, WRF-Sfire fire plume behavior is consistent with the interpretation of FireFlux observations. The study suggests optimal experimental pre-planning, design, and execution of future field campaigns that are needed for further coupled atmosphere-fire model development and evaluation.

# 1. Introduction

Over the last two decades, significant advances have been made on the development of coupled atmosphere-fire numerical models for simulating wildland fire behavior. While numerical studies using coupled atmosphere-fire models have shed light on the dynamics of fire-atmosphere interactions (Clark et al. 1996, Morvan and Dupuy 2001, Linn et al. 2002, Linn and Cuningham 2005, Coen 2005, Cunningham et al. 2005, Sun et al. 2006, Mell et al. 2007, Cunningham and Linn 2007), none of these models have been evaluated or validated using in-situ, field-scale observational data. This is due to the lack of field measurements appropriate for model testing. The objective of this study is to determine the ability of the WRF-Sfire modeling system (Mandel et al. 2009, 2011) to predict observable phenomena accurately by comparing model output to comprehensive field measurements. Measurements made during the FireFlux field experiment (Clements et al 2007, 2008; Clements 2010) are used for this purpose.

No single numerical wildfire behavior prediction model available today is ideal. Existing wildfire behavior prediction models range from the mainly physical, based on fundamental understanding of the physics and chemistry involved, to the purely empirical, based on phenomenological descriptions or statistical regressions of fire behavior. As a result, these models differ greatly in terms of physical complexity, representation of atmosphere-fire coupling, extent of resolved versus parametrized processes, and computational requirements. For both research and operational use, each has its strengths and weaknesses. A major difficulty in developing realistic wildfire behavior prediction models is the lack of observational data in the immediate environment of wildland fires that can be used for validating these models (Clements et al 2007).

WFDS (Wildland Urban-Interface Fire Dynamics Simulator; Mell et al 2007) and FIRETEC (Linn 1997; Linn et al. 2002) are two examples of the most advanced fire-scale coupled fire-atmosphere wildfire behavior models. This class of model attempts to represent localized fire–atmosphere interactions with explicit treatment of convective and radiative heat transfer processes. Computational resources are dedicated to resolving the fine-scale physics of flame, combustion, radiation, and associated convection. Unfortunately, the computational demands of these models preclude their use as operational field models for wildfire behavior forecasts. Using current computer technology, the wall-clock time required to complete a wildfire simulation contained in even small-sized (e.g., *x,y,z* dimensions less than 4 km x 4 km x 2 m) domains is significantly greater than the simulated fire's lifespan; by the time the forecast is computed, it is already outdated. Furthermore the small domain size generates often non-physical numerical boundary effects (Mell et al. 2007). Typically run as a stand-alone model in research mode, wildfire simulations by these models lack a real-time multi-scale Atmospheric Boundary Layer (ABL) wind and weather forecast component.

At the other end of the model spectrum are the current operational real-time wildfire behavior prediction models (Sullivan 2009; Papadopoulos and Pavlidou 2011). These are the simplest models that, instead of solving the governing fluid dynamical equations, rely on semi-empirical or empirical relations to provide a fire's rate of spread as a function of prescribed fuel properties, surface wind speed and humidity, and a single terrain slope. The main advantages of these models are that they are computationally very fast and can be run easily on a single laptop computer. The main disadvantage is that they are limited physically. These models consider only surface wind direction and strength, they lack a real-time multi-scale wind and weather forecast component, and they cannot account for coupled atmosphere/wildfire interactions. The implication is that these models perform well for cases when atmosphere-fire coupling provides for steady-state fire propagation under environmental wind conditions stable to flow perturbations. Applications of these empirical and semi-empirical models to wildfire conditions where fire–atmosphere coupling does not provide for steady-state propagation (e.g., crown or high intensity fires, or wildfires in complex terrain or changing environmental wind conditions) can lead to serious errors in fire spread predictions (Beer 1991, Finney 1998).

There exists an intermediate class of wildfire behavior prediction model that may be categorized as a "quasi" physical coupled atmosphere-fire model (Sullivan 2009). This class of model includes the physics of the coupled fire/atmosphere but obtains heat and moisture release rates, fuel consumption, and fire-spread rate from the same prescribed formulae or semi-empirical relations that are employed by current operational fire behavior models. Based on operational fire-spread formulations driven by the coupled fire-atmosphere winds at the fire line, a simple numerical scheme is used to move the fire perimeter through the fuel and each surface model grid. Computational resources are therefore dedicated to resolving the atmospheric physics and fluid dynamics at the scale of the fire line. The highly simplified treatment of combustion, radiation, heat transfer, and surface fire spread makes these models perform significantly faster than physics-based ones, and therefore these models appear to be good candidates for future operational tools for wildfire forecasting.

Examples of this type of model are CAWFE (Coupled Atmosphere-Wildland Fire-Environment; Clark et al. 1996, 2004, Coen 2005), fire-atmosphere coupled UU LES (University of Utah Large Eddy Simulator; e.g. Sun et al. 2009), MesoNH-ForeFire (Filippi et al. 2009), and WRF-Sfire (Mandel et al. 2009, 2011). Even though atmospheric and fire components differ, these models are based on the same operating principles (Sullivan 2009). Proponents of these models argue that, if the goal is a real-time operational physically-based fire behavior forecast model, then this approach is feasible provided the sub-grid scale parametrizations of fire produce accurate heat release rates, and the mathematical algorithms propagating the fire at rates specified by the empirical fire-spread formulations calculate realistic spread rates under coupled fire-atmosphere wind conditions. Of these models, only WRF-Sfire has access to a real-time multi-scale forecast of ABL flow, making it the most appropriate candidate for operational wildfire prediction.

This study attempts, therefore, to determine the ability of the WRF-Sfire modeling system to predict accurately observable phenomena by comparing model output to comprehensive field measurements. WRF-Sfire prediction is evaluated from the point of view of fire-atmosphere interactions, and in situ measurements collected at the fire line during the FireFlux experiment (Clements et al 2007) are employed for this purpose. FireFlux's fire line wind and temperature measurements are used to evaluate and improve WRF-Sfire fire line's predicted ROS (Rate-of-Spread), temperatures, and winds. The uniqueness of FireFlux compared to the open grassland fire experiments conducted in Australia (Cheney et al 1993; Cheney and Gould 1995) is that it recorded the state of the atmosphere during the FFP (Fire Front Passage), rather than focus on fire line depth and spread. When comparisons between observations and WRF-Sfire predictions indicated good agreement, the simulation was used to display the flow features observed during FireFlux in terms of WRF-Sfire predicted fire spread, plume properties, and behavior.

The paper is organized as follows. Section 2 describes the field experiment used for the WRF-Sfire model validation. The model description and its setup are described in Sections 3 and 4. Results on fire spread, and thermal and dynamical plume properties, and the structure of the fire-induced flow are presented in Section 5 and compared to FireFlux observations. In Section 6, adjustments made to WRF-Sfire to obtain the agreement with FireFlux observations are discussed, and suggestions are made for the design of future field campaigns to deliver the observations necessary for evaluation or validation of existing coupled atmosphere-fire prediction models. Concluding remarks are given in Section 7.

## 2. Overview of the FireFlux Experiment

The FireFlux experiment took place on 23 February 2006 at the Houston Coastal Center, a 1000-acre research facility of the University of Houston. The FireFlux experiment is the most intensively instrumented grass fire to date. The experiment was designed to study fire-atmosphere interactions during a fast-moving head fire in grass fuels by measuring the wind, turbulence, and thermodynamic fields of the near-surface environment and of the plume. An overview of the experimental design, and results of the turbulence and thermodynamic measurements are found in Clements et al. (2007, 2008) and Clements (2010), respectively.

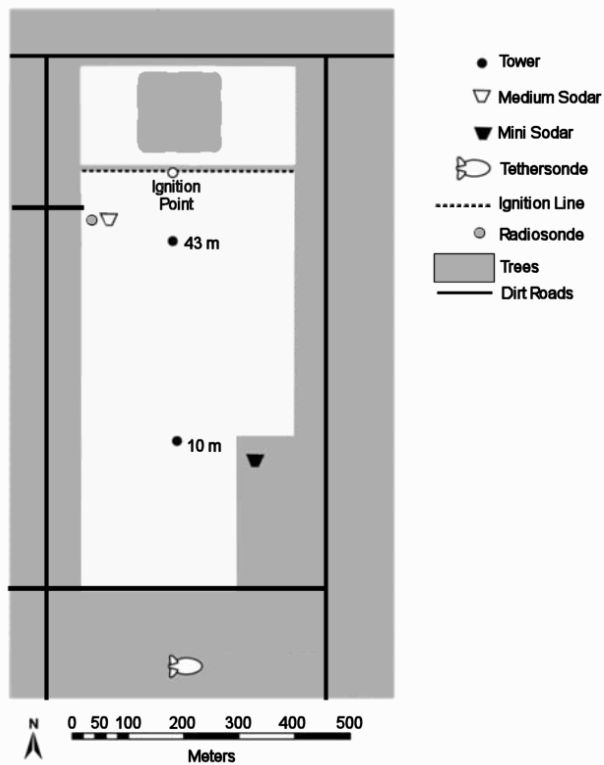

Fig. 1. Instrument locations and the layout of the FireFlux experiment. White area indicates grass.

Figure 1 shows the experimental layout with instrument locations. The key platforms included a multi-level 43 m micrometeorological flux tower located in the middle of the fuel bed and a similarly instrumented, but shorter, 10 m tower located 300 m downwind from the 43 m main tower. These two towers are hereafter refereed to as MT (for main tower) and ST (for short tower). In addition to MT and ST, a tethered balloon system was deployed on the downwind edge of the burn block to measure temperature, humidity, and wind speed and direction at five altitudes up to 150 m Above Ground Level (AGL). Two SODARS were also used; one was a medium-range system located on the northwest corner of the fuel bed, and the other a mini-sodar located at the southeastern corner of the burn block. Additionally, a radiosonde was released at the edge of the burn block, providing a full in-situ vertical sounding of temperature, humidity, wind speed and wind direction. Video and time-lapse photography were used to record fire behavior and the spread rate of the fire front.

## 3. Model description

WRF-Sfire (Mandel et al. 2009, 2011a,b) combines the Weather Research and Forecasting Model (WRF) with a semi-empirical or empirical fire spread model. The fire model runs on a refined mesh at surface level. In each model time step, the near-surface wind from WRF is interpolated vertically to a logarithmic profile and horizontally to the fire mesh to obtain height-specific wind that is input into the user-chosen fire spread-rate formula. In this study the Rothermel fire spread-rate formula (Rothermel 1972) was used to determine, based on the fuel properties and WRF wind speed, the instantaneous fire spread rate at every refined mesh point. Fire propagation is implemented on the fire mesh by the level-set method (Osher and Fedkiw 2003) and applying Rothermel's fire spread formula in the direction normal to the fire line. After ignition, the fuel is set to decay exponentially with time, with the time constant dependent on fuel properties. The latent and sensible heat fluxes from the fuel burned during the time step are averaged over the cell of the atmosphere model and inserted into the lowest levels of the atmospheric model, assuming exponential decay of the heat flux with height. Fuels are given as one of 13 categories (Anderson 1982), and associated with each category are prescribed fuel properties such fuel mass, depth, density, surface-to-volume ratio, moisture of extinction, and mineral content. The model supports point, instantaneous line, and "walking" ignitions. The Sfire model is embedded into the WRF modeling framework enabling easy set up of idealized cases or real cases requiring realistic meteorological forcing and detailed description of the fuel types and topography. The nesting capabilities of WRF (not used in this study) allow for running the model in multi-scale configurations, where the outer domain, set at relatively low resolution, resolves the large-scale synoptic flow, while the gradually increasing resolution of the inner domains allows for realistic representation of smaller and smaller scales, required for realistic rendering of the fire convection and behavior. The Sfire model is available from openwfm.org; a limited version is available in WRF release since version 3.2 as WRF-Fire.

## 4. Model setup

The WRF modeling framework is used for routine numerical weather prediction in the United States, and its incorporation in WRF-Sfire allows for detailed descriptions of the land use and fuel types (Beezley 2011, Beezley et al. 2011). In this study, these capabilities were extended to the use of standard land surface models, custom topography, and land use and fuel categories (defined in external files), without the need of the WRF preprocessing system. The aerial picture of the experimental site, model domain boundary, land use, fuel map, and ignition line are presented in Fig. 2.

The fuel map used in the WRF-Sfire FireFlux simulation was initialized with the map of land use derived from an aerial Google Earth picture and simplified to two USGS land use types: mixed forest and grassland. The grass fuel was designated as tall grass fuel, category 3, and the surrounding area as noncombustible fuel, category 14 (Anderson 1982). More details about the fuel characteristics are given in Table 1. Model surface properties defaulted to either one of these two fuel categories. The grass roughness length was determined to be 0.02 m according to the pre-fire wind profile measurements from the FireFlux experiment.

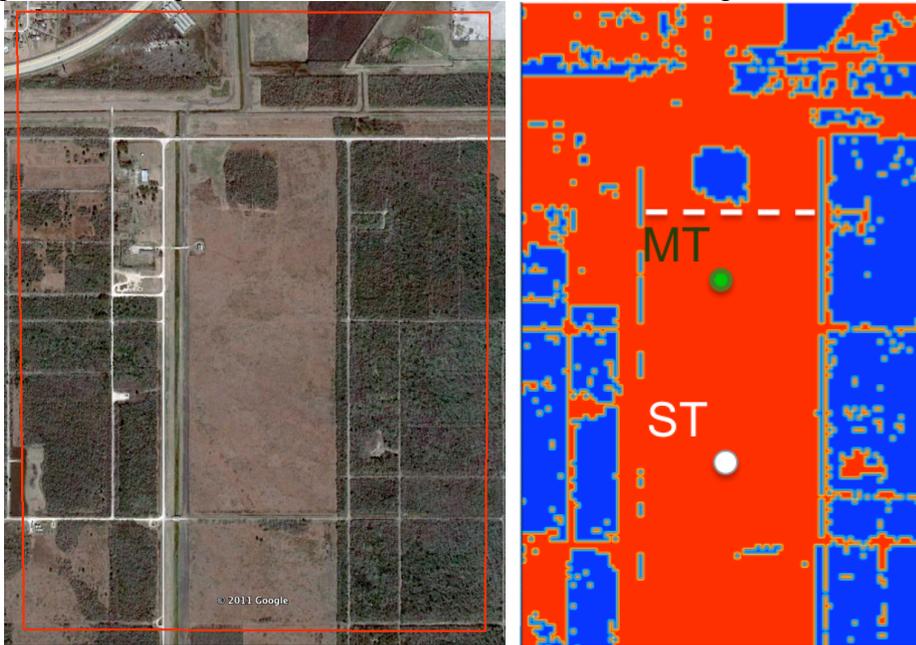

Fig. 2. a) Aerial picture of the FireFlux area with the domain boundary marked in red; b) "Land use" field from WRF input (red signifies grassland, blue signifies mixed forest) with locations of main tower (MT; green dot), short tower (ST; white dot), and ignition line (white dashed line).

The ($x,y,z$) dimensions of the model domain are (1000m,1600m,1200m). The WRF atmospheric computations were performed on a regular horizontal grid of 10 m spacing and of non-uniform vertical-grid spacing, stretched using a hyperbolic function, varying from 2m at the surface to almost 34m at model top. The fire model mesh was 20 times finer than the atmospheric x,y mesh, what translates into a 0.5m horizontal grid spacing. The computational details are presented in Table 1.

Thermocouple measurements at 0.13m AGL reported a uniform fire domain temperature of 19.22C before ignition, and this value was used as the model's initial surface temperature. Initial wind, temperature, and moisture fields were reconstructed using vertical profiles taken from the MT measurements up to 43m AGL, the tethersonde measurements for 43-130m AGL, and the morning sounding measurements for 130-1200m AGL. The initial model profiles for wind speed and direction, and potential temperature are seen in Figure 3. The atmosphere was slightly unstable for the first 50m AGL due to intense solar heating of the surface, and neutral above and up to approximately 400m AGL. The wind was northerly at 3 m s$^{-1}$ for the first 2m AGL, and increasing in magnitude with height to approximately 7 m s$^{-1}$ at 50m AGL, becoming more north-northwesterly. At higher levels, up to 400m, wind speed was fairly uniform, averaging about 8 m s$^{-1}$. There was a marked deviation in wind speed and direction at approximately 50m AGL. The reason for this is unknown, assumed to be an artifact of combining tower and tethersonde data, and this deviation was not removed from the data set.

WRF-Sfire's "walking ignition" option was used to emulate the start of the fire. Fire line ignition started at the approximate center of the burn area (see Fig. 1) and progressed sideways at the speed estimated by GPS data collected during the actual ignition procedure. Since the GPS unit recorded only one ignition branch, the timing of the other branch was based on data collected during a walk along the ignition line after the actual ignition procedure. The overall length of the ignition line was 385m. The ignition procedure took around 2.5 minutes and the whole burn took around 17 minutes. More details on the ignition procedure are given in Table 1.

In previous versions of WRF-Sfire, a point ignition was modeled by setting a fixed circle on fire at once, with the circle size at least the size of the horizontal fire cell, while a walking ignition was modeled as a succession of circles. In this study, such a walking-ignition scheme produced an ignition line at least 0.5 m wide, while FireFlux's dip torch ignition line was likely thinner; the 0.5m-wide ignition strip caused the initial fire propagation to be too fast. Therefore, to prevent this, WRF-Sfire ignition model was revised to apply a slower initial sub-grid ROS to the time period from ignition until the fire is large enough to be visible on the fire mesh, after which time the propagation mechanism based on the Rothermel formulation takes over. See Mandel et al. (2011), their Section 3.6, for the details of the ignition implementation in the framework of the level set method.

In addition, to achieve a realistic fire propagation rate between ignition of the initial fire line and FFP at the MT, the Rothermel default no-wind fire line Rate-of-Spread (ROS) was increased from 0.02 m s$^{-1}$ to 0.1 m s$^{-1}$ (Table 1). This ROS is applied when there is no wind component perpendicular to the leading edge of the sub-grid-scale combustion zone. Comparison with flank ROS simulated by FIRETEC (Cunningham and Linn 2007) for grass fires suggests that 0.02 m s$^{-1}$ is an order of magnitude too small. The five-fold increase in no-wind ROS also resulted in more realistic spread along the fire's flanks and back.

Another fire model feature that was set to provide good agreement with observations was the e-folding extinction depth used to parametrize the absorption of sensible, latent, and radiant heat from the fire's combustion into the surface layers of WRF. In WRF-Sfire, the total heat liberated into the atmosphere by the fire is released into the vertical model atmosphere using the e-folding extinction depth. Sun et al (2006, 2009), following Clark et al (1996a), also used this simple extinction depth approach to treat the fire-atmosphere heat exchange. Sun et al (2006) found that plume-averaged properties were sensitive to the choice of extinction depth; too large an extinction depth under estimated important near-surface properties just above the combustion zone, such as temperature excess and vertical plume velocity; two small an extinction depth produced agreement between observed and model predicted plume-averaged temperatures, but less agreement between observed and model-predicted plume-averaged vertical velocities, just above the surface. There exists therefore no set value for this parameter. In this study the flame length estimate of 5.1 m by Clements et al (2007) was used to set the extinction depth to 6 m.

Unfortunately, the infrared camera used to video the fire experienced technical problems and continuous infrared imagery of the location and spread rate of the fire head is not available for analysis. Wind and air temperature measurements are used instead to represent head fire spread, plume properties, and behavior. Note that the FireFlux temperatures used in this study were measured by a Type-T 1Hz thermocouple (Clements et al 2007; their Table 1). FireFlux temperatures were also measured at 2.1m AGL at the MT and 2, 5, and 10m AGL at the ST with a type-K fine-wire 20Hz thermocouple. Because the fire-wire thermocouples failed at times, and measurements below 2.5m were possibly affected by precautions taken to shield these thermocouples from damage by the fire, these data are not used in the evaluation of WRF-Sfire output.

Horizontal atmospheric grid resolution limits the frequency of the fluctuations in temperatures or flow that the model can resolve. For the atmospheric horizontal grid size of 10m, the shortest disturbance or fluctuation that the model resolves is assumed to have an approximate length of 40m. If this perturbation travels 8 m s$^{-1}$, roughly the peak wind speed observed during FireFlux, the effective frequency of disturbance resolved by the model is 1/(8/40) or 0.2Hz. Therefore, the WRF-Sfire output frequency was 0.2Hz, results were saved every 5 seconds, and a 5s moving average was applied to the FireFlux measurements for direct comparison to model results. Simulated plume temperatures and winds represent integrated values over relatively large model grid volumes and time steps, and are not expected to depict variance as large as the variance in local instantaneous temperature or wind values measured during FireFlux.

## 5. Results

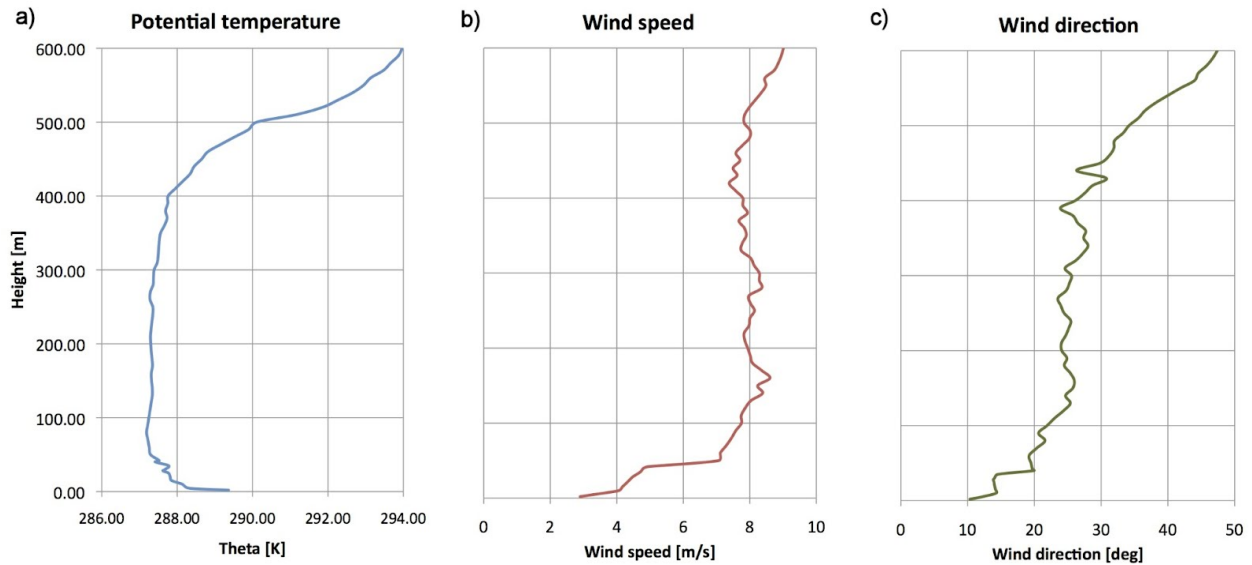

Fig. 3. Initial atmospheric profiles used for model initialization: a) potential temperature; b) wind speed; and c) wind direction.

*5.1 Fire spread*

Fire spread rates are determined by time series of 4.5m MT and 5m ST AGL simulated and observed air temperatures shown in Figure 4 (grey lines show 1Hz thermocouple data and black lines show 5s averaged 1Hz thermocouple data). Model results are interpolated vertically between second (4.49m) and third (7.7m) model levels. The timing of FireFlux's FFP through the MT is indicated by rapidly rising and falling air temperatures in the time series, and this timing is well captured by WRF-Sfire. The simulated MT air temperature reached the peak value at 225s from the ignition, while observations indicate a peak temperature just 6s earlier. Timing of the FFP through the ST is also well captured by the model. There is only a 5s delay with respect to the observations, and the simulated ROS between the two towers is 1.61 m s$^{-1}$, exactly the observed ROS.

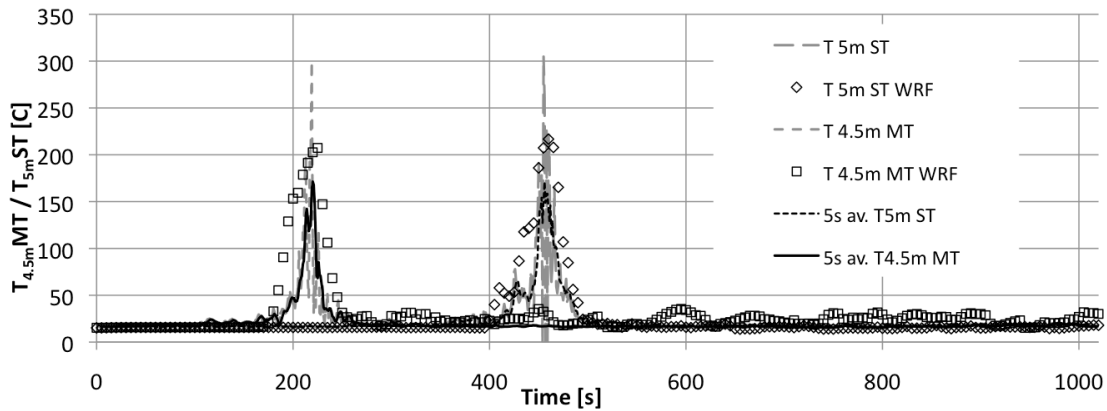

Fig. 4. Time series of the 4.5m AGL air temperature at the location of the main tower (MT) and 5m AGL air temperature at the short tower (ST). Gray lines show 1Hz measurements, black lines show 5s averaged values, and symbols (diamond and square) show model data.

In terms of magnitude, the agreement between observed and simulated temperatures is relatively good. Figure 4 indicates that the WRF-Sfire's peak air temperature at the MT is 35 degrees (20%) warmer than the 5s-averaged measurements and 88 degrees cooler than the maximum temperature from the 1Hz thermocouple data. Compared to the Type-K fine-wire thermocouple, used to measure burning grass fuel temperature (sampling frequency 20Hz; see Clements et al 2007, their Table 1), the data from the ordinary thermocouple, used to measure air temperature (sampling frequency 1Hz), show that at 4.5m AGL tended to underestimate air temperatures by as much as 90C and by 32C after 5s averaging. This suggests that simulated air temperatures are within only 3C of temperatures measured with a faster responding fine-wire thermocouple. Figure 4 shows that ST thermocouple temperatures are slightly higher than those at MT. Temperature maxima are 304C at the ST and 295C at the MT. The simulated peak temperature at the ST is also 9C higher than the simulated peak temperature at the MT. These differences are eliminated by 5s averaging, and filtered peak air temperature is 172C at the MT and 171C at the ST. The model underestimated again the 4.5m AGL air temperature at the ST by 88 degrees, almost exactly the bias between model and 1Hz temperature data at MT. Compared with the filtered data, the model overestimated the ST air temperature by 45 degrees.

*5.2 Thermal plume structure*

Figure 5 is the same as Figure 4 except for time series plots at the MT at 2m, 10m, 28m and 43m AGL, and demonstrates how well WRF-Sfire plume's vertical temperature profile matches tower thermocouple temperature measurements. Tower temperatures before and after fire passage remain steady, deviate very little from the background temperature, and this behavior is well predicted by WRF-Sfire. Figures 5 a) and b) show that temperatures in the WRF-Sfire plume begin to rise above and fall to ambient (no fire) values at virtually the same times as FireFlux plume values; i.e., fire-plume arrival and passage are practically identical for both measured and simulated plumes. However, changes in observed temperature with fire passage do differ from model results. FireFlux temperatures rise slightly just ahead of a rapid increase to peak temperature values, while model temperatures do not show a strong tendency towards "preheating" and generally begin a more immediate but less abrupt rise. While FireFlux temperatures peak, decline abruptly, and then decay away to almost ambient values as the fire passes, the smooth fall in WRF-Sfire temperatures after the peak matches generally the smooth rise in temperatures before the peak. At higher elevations (Figure 5 c and d), the WRF-Sfire plume temperatures rise on average at almost the same rate, but fall sooner, than the FireFlux temperatures. This temporal shift may be attributed to either a slight underestimation in the simulated horizontal plume extent at higher elevations or that the fine-scale fire plume structure is unresolved in the WRF-Sfire simulation. The generally slow rise and fall in simulated temperatures may be the consequence of either the coarse model output time interval (5s), the relatively course atmospheric grid volume over which model variables are averaged, or that the fire model burns all fuel and releases heat at the same rate. These model oversimplifications may also be responsible for the unrealistic lack of spatial and temporal temperature (and wind) fluctuations in the WRF-Sfire plume, especially at levels > 10m AGL. Differences between model and fireflux thermocouple temperatures are to a great degree eliminated by 5s averaging. When WRF-Sfire temperature time series in Figures 4, and 5 a), b) are compared to the 5-point moving mean of the FireFlux temperatures, a greater level of agreement is seen, where, to a moderate degree, WRF-Sfire under predicts fire plume temperatures (by 20%) at 4.5 m but agrees within 15% at all other levels.

a) Temperature 2.1m main tower (fuel)

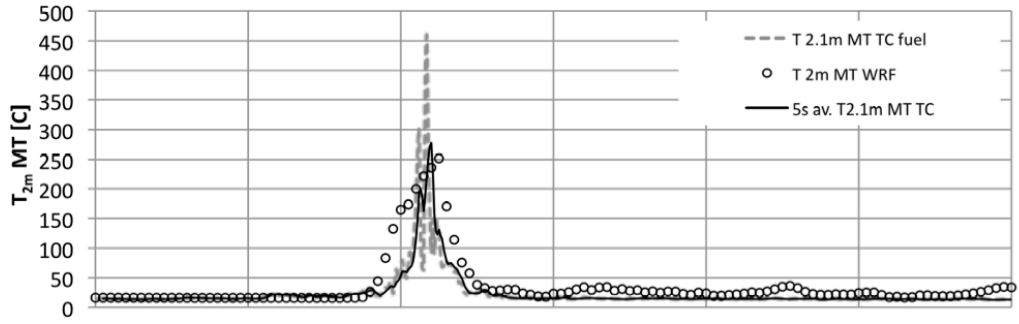

b) Temperature 10m main tower

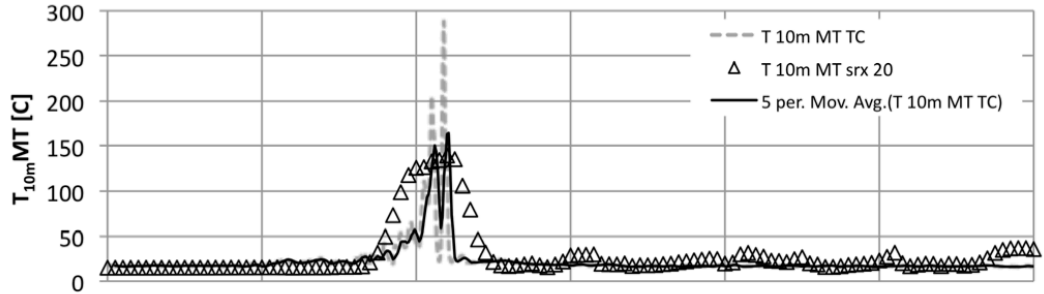

c) Temperature 28m main tower

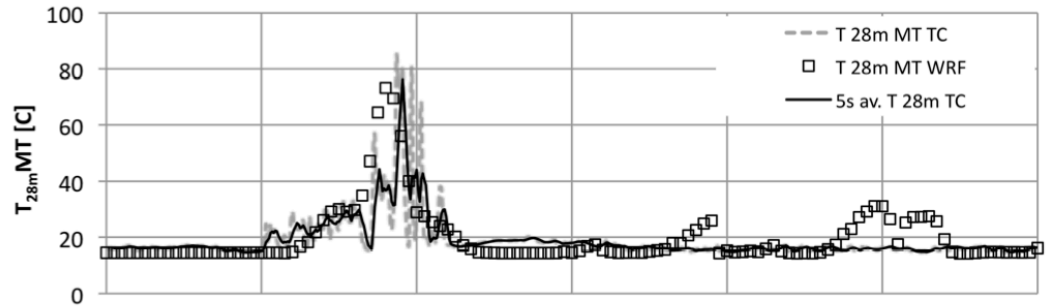

d) Temperature 42m main tower

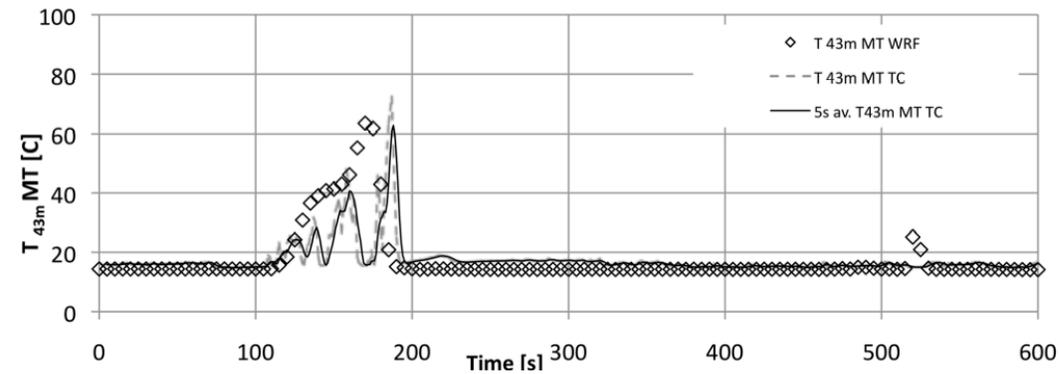

Figure 5. Time series of the thermocouple air temperatures at the location of the MT at: a) 2.1m, b) 10m, c) 28m, and d) 42m AGL. Gray lines show 1Hz measurements, black lines show 5s averaged values, symbols (open circles, triangles, squares, and diamonds) show model data.

Figures 5 c) and d) show the upper levels of the warm, downwind-tilted FireFlux plume arriving, respectively, at the main tower just at and after 100 s into the experiment. Plume arrival occurs slightly sooner at 28 m AGL compared to 43 m AGL, and plume passage occurs later at 28 m AGL compared to 43 m AGL. Although WRF-Sfire temperature time series in Figures 5 c) and d) do not show plume arrival at lower levels first, the temporal differences in fire-plume arrival and passage between FireFlux and WRF-Sfire at these AGLs are slight. Measured plume temperatures as well as the 5-point moving means during fire passage show significant fluctuations in magnitude at both 28 m and 43 m AGL. Fluctuations of this magnitude are not unexpected in the upper portion of an entraining, turbulent fire plume. The results indicate that even though the WRF-Sfire did not capture these high-frequency fluctuations, it predicted the FireFlux peak temperatures at 28 m and 43 m AGL very accurately (with -4.6% and +1.6% bias, respectively). Time of plume arrival is well predicted by WRF-Sfire at the 43 m level and under predicted by approximately 20 s at the 28 m level. The abrupt fall-off in measured plume temperatures as the upwind edge of the plume passes the tower is well represented in the WRF-Sfire time series. Temperature measurements at 43m show that air temperatures remain slightly elevated above ambient values even after the plume has passed, while temperatures measured just one meter below (not shown) and simulated by WRF-Sfire drop immediately to pre-fire ambient values. However local variation of plume properties in the upper-levels of a highly turbulent convective plume is not unrealistic, which suggests that this level of agreement between predicted results and measurements remarkable. Clements (2010) reports that the greatest temperatures difference and variability compared to ambient air temperatures occurred at 10 m AGL, where entrainment of ambient air is possibly the greatest.

Figure 6 is the same as Figure 5 except for time series plots at the ST at 2m and 10m AGL. Fire-plume arrival and passage are practically identical for both measured and simulated plumes. However, WRF-Sfire overestimates plume temperatures at these two levels. Simulated fire-plume temperatures are within 10% of the 1Hz observations, but greater by 82C at 2m AGL and 45C at 10m AGL than peak 5-point moving means, and they remain elevated for a significantly longer time than measured ones. Due to the lack of infrared video camera recordings, it is difficult to report the actual fire front depth. However, differences in the time periods between simulated and observed fire plume temperature values suggest that the model is overestimating the thickness of the fire front. Using a 100kW m$^{-2}$ heat release rate threshold, the simulated fire front thickness at the ST is estimated at 45m, which appears to be too large. Note that at the MT, the fire front thickness is estimated to be half as large, only 27.5m thick. This 45m front thickness is likely responsible for an unreasonably higher fire heat release and consequently unrealistically higher model fire-plume temperatures at the ST .

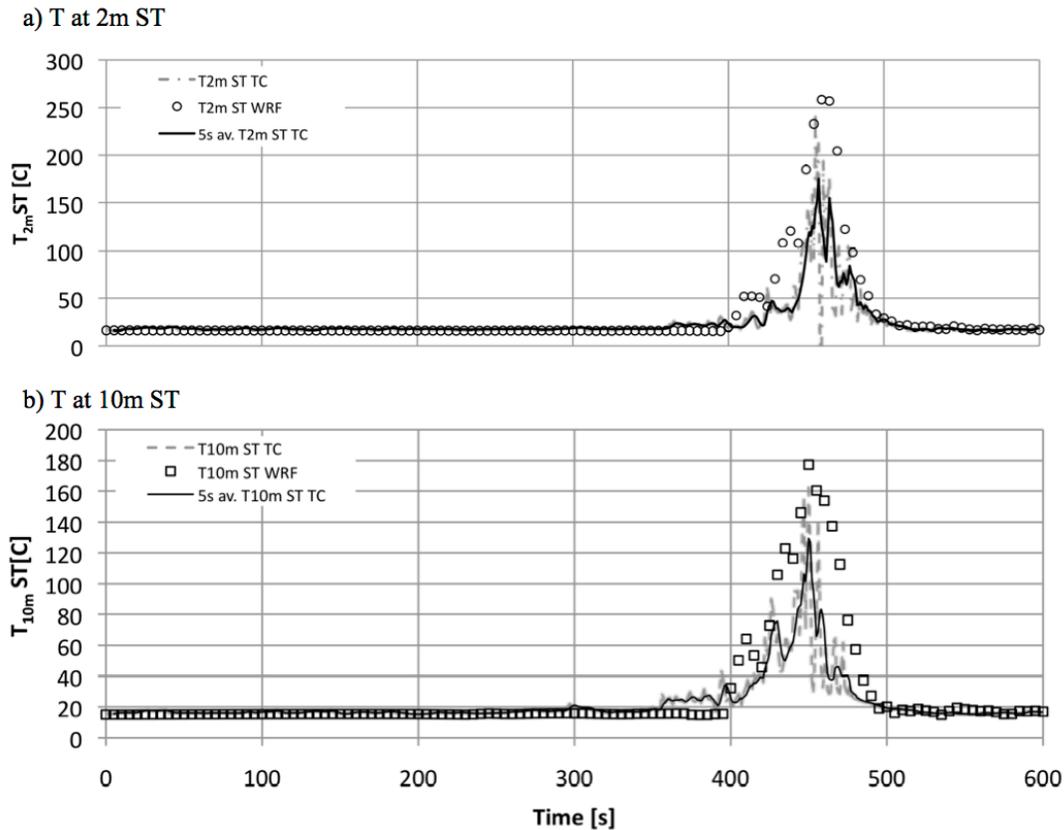

Figure 6. Time series of the fine thermocouple air temperatures at the location of the short tower (ST) at a) 2m, b) 10m. Gray dashed lines show 1Hz measurements, black solid lines show 5s averaged values, and symbols (open circles and squares) show model data.

Figure 7 shows plots of contoured WRF-simulated (upper plot) and thermocouple measured (middle and lower plots) temperatures at the MT as a function of time. Figures 7c and 7b show that heating by the FireFlux fire front and passage is rapid and limited to a small volume (below 15m AGL) around the combustion zone as the fire front quickly propagates downstream. Owing to entrainment and turbulent convection in the plume, FireFlux temperatures display a large degree of variance (Clements *et al.* 2008; Clements 2010). The averaged measured temperature maximum starts around 210s and lasts till 220s since ignition (Figure 7b). That implies that the fire thickness computed based on the average rate of spread between the towers was probably no greater than 6.2m (10s/1.61ms$^{-1}$). The simulated temperature maximum starts at similar time, but lasts significantly longer (till 235s), indicating that the thickness of the simulated fire front was at least three times wider than the observed one. As discussed in Section 4, the horizontal resolution of the atmospheric model directly controls the minimum width of the temperature perturbation that can be resolved. The averaging of heat released by the fire over the whole atmospheric grid-volume affects the appearance of the fire signal on the atmospheric mesh. Regardless of how narrow the fire front on the fine fire mesh is, as the fire crosses two adjacent atmospheric cells, the heat released gets averaged over the two cells. As a consequence the minimum width of the fire-related thermal signal seen on the atmospheric grid is two atmospheric grid spaces, which in this study is 20m, far greater than the estimated 6.2m fire-front thickness. The fuel burn rate used in WRF-Sfire is the same for all fuel types, which may result in a too long fuel residence time for quickly burning fuels like grass. That may also result in the overestimation in the width of the fire zone as evident in Figure 7 a).

a)
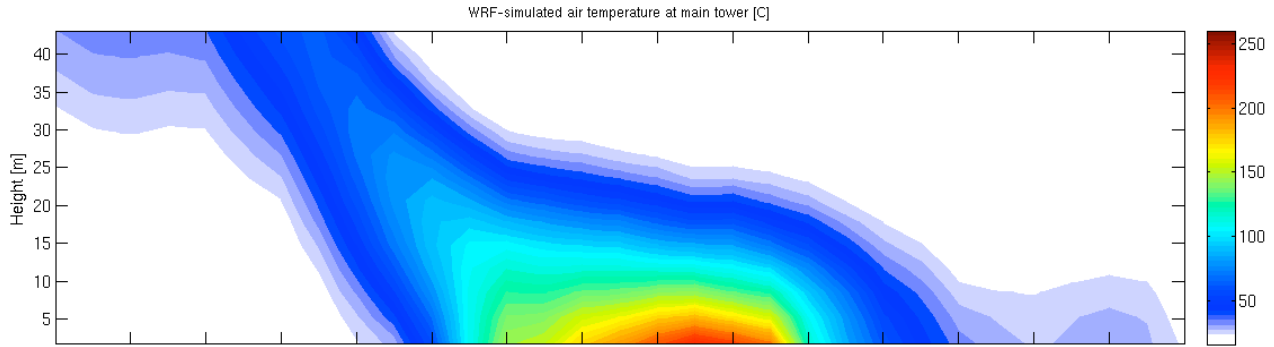

b)
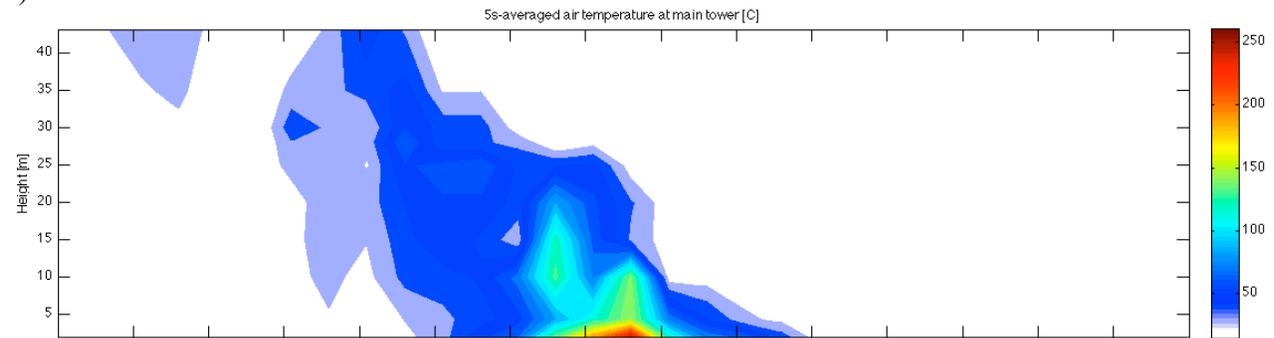

c)
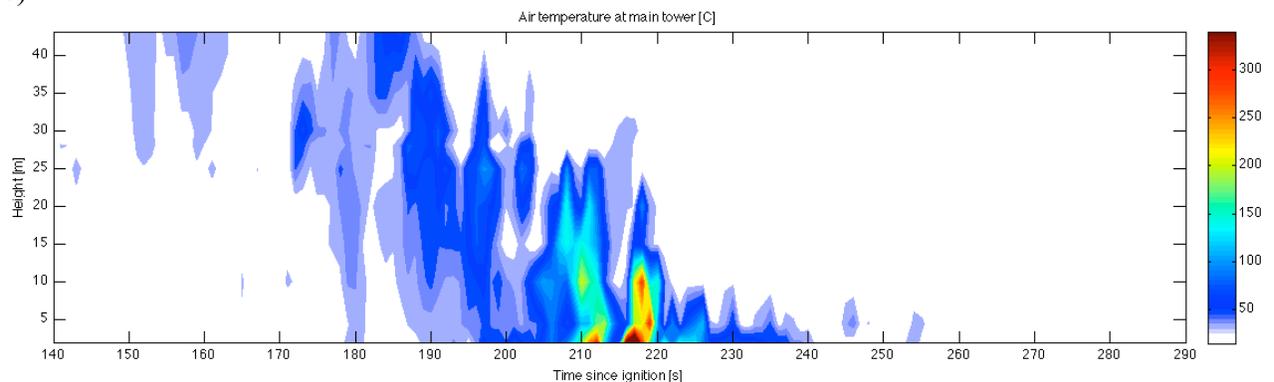

Figure 7. Air temperatures at the MT as a function of time: a) WRF-simulated, b) thermocouple 5s averaged, c) thermocouple raw (1s).

Nonetheless, Figure 7 shows that WRF-Sfire successfully captured the plume's downstream tilt, the arrival between 180 to 200s of fire-warmed surface air, and the passage of the fire-warmed surface air at approximately 260s, with the low-level near-surface warmest volume of air arriving approximately 10s later at the MT than observed. Contoured WRF results also show that the 15m vertical extent of the warmest (greater than 100C) plume temperatures matches the observations presented in Figure 7 b).

*5.3 Dynamical plume structure*

*5.3.1. Fire-induced horizontal winds*

WRF-Sfire computes the ROS based on coupled fire-atmosphere winds at the fire line. It is crucial, therefore, for realistic prediction of wildfire behavior that WRF-Sfire capture accurately the fire-atmosphere interaction and evolution of the surface flow at the fire line. To evaluate for this, model results are compared to FireFlux wind measurements. Heat and temperature extremes did cause some minor damage and instrument failure during FireFlux. The sonic anemometer at the ST broke during the FFP. Therefore in the analysis of the WRF-Sfire plume dynamics, data from the MT, which captured more of the vertical plume structure, are used.

The time series plots of the wind speed measured by the sonic anemometer (dashed line) and simulated by WRF-Sfire (symbols) at the MT at AGL levels 2m, 10m, 23m, and 43m are shown in Figure 8. The solid lines are the 5-point moving means of wind speed measurements.

The FireFlux time series in Figure 8 show disturbed wind speeds before, during, and after the fire plume passes the MT. Passage is not marked by a distinct rise and fall in wind speed as it was with temperature, and this is especially true at upper-tower levels 28m (Figure 8 c) and 43m (Figure 8 d). At 2m AGL (Figure 8 a) just before fire passage wind speeds rise, reaching 6 to 12 m s$^{-1}$ during fire passage, and then fall to values slightly greater than ambient just after fire passage. Wind speeds at 10m AGL (Figure 8 b) show similar behavior except that peak values are lower, approximately 4 to 8 m s$^{-1}$. Both measured and 5-point moving means in Figure 8 c) and d) show strong fluctuations in wind speed as the FireFlux plume passes the MT. At these levels the FireFlux measurements vary in magnitude and do not display a single peak value.

There is agreement in Figure 8 between the WRF-Sfire results and the FireFlux 5-point moving means. Figures 8 a) and b) show how, during fire passage, although wind speeds fluctuate throughout, the overall trend is well captured by WRF-Sfire. Simulated and observed wind speeds rise, peak, and then fall. At 10m AGL the maximum simulated wind speed matches almost exactly the filtered observations (with a 0.2 m s$^{-1}$ negative bias), while at 2m AGL the model overestimates the peak wind speed by only 1.2 m s$^{-1}$. Neither time series of observed or model wind speeds at the higher elevations display a strong response to the fire plume's passage. At these levels fluctuations in ambient wind speed are similar in amplitude to those associated with fire plume passage, making the quantification of the fire's effect on the wind speed practically impossible. It can be said that before plume passage WRF-Sfire wind speeds at 28m and 42m are in overall mean agreement with FireFlux observations. After plume passage, WRF-Sfire wind speeds at 28m and 42m are overall greater than FireFlux observations. As discussed before, considerable variation of plume properties in the upper-levels of a highly turbulent convective plume is not unrealistic, which makes even this level of agreement between predicted results and measurements acceptable.

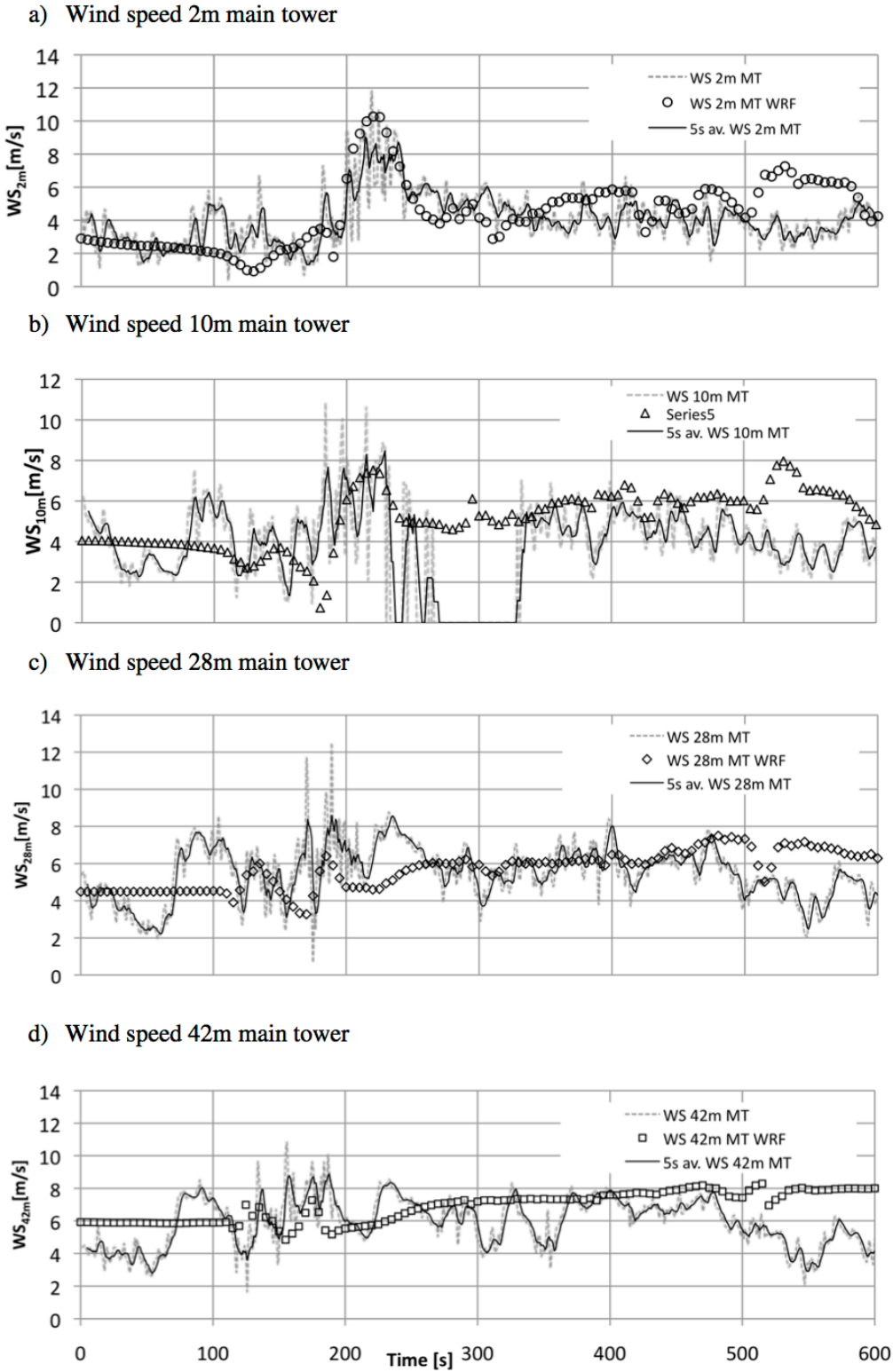

Figure 8. Time series of horizontal wind speed (WS) at MT levels: a) 2m, b) 10m, c) 28m, and d) 42m. Gray dashed lines show 1Hz measurements, black solid lines show 5s-averaged values, symbols (circle, triangle, diamond, square) show model data at the four MT measurement levels.

The WRF-Sfire wind speeds shown in Figure 8 behave as described by Clements et al (2007). As the fire front approaches the MT the surface wind speed more than triples, and before the horizontal wind increase, there is a brief period of calm that as suggested by Clements et al (2007) is associated with horizontal convergence in the flow ahead of the fire line that coincided with increased vertical motion. Clements et al (2007) has the wind direction shifting from northeasterly to southerly at 12:45:50 CST, approximately 50 s before the head fire reached the MT. As the fire front passed the MT at 12:46:40 CST, wind direction switched back to ambient northerly flow, while wind speeds increased from approximately 3 m s$^{-1}$ to over 10 m s$^{-1}$. At the upper levels of the MT, there were large increases in wind speed, but not as long in duration as observed at the surface. While the vertical profile of the ambient wind shows wind speed increasing almost logarithmically with height, both observations and the simulation indicate that, during passage of the fire front, the maximum wind speed occurs at the surface and decreases in magnitude with height.

*5.3.2. Fire-induced updraft*

Figure 9 is the same as Figure 8 except for vertical wind speed. The first fire-induced updraft occurs roughly 200s into the simulation, as the fire line approaches the MT, and Figure 8a shows that this occurs around 25 s before the peak in temperature. The updraft passes the tower, and is then followed by a strong downdraft. Figures 7a and 8a suggest that, at 2m AGL, the updraft is not collocated with the maximum horizontal wind speed as originally suggested by Clements et al (2007). The model's ability to resolve the updraft velocity at 2m AGL is limited. The 2m height corresponds roughly to the model's first AGL level. Since vertical velocity is set to zero at the first model level (ground), the model underestimates vertical wind variations close to the surface. Nonetheless, as shown in Figure 9a, at 2m AGL, the updraft followed by a decrease in the vertical velocity and downdraft of similar strength are still captured realistically by the model.

The model and FireFlux observations displayed in Figure 9 show that the maximum updraft velocity associated with plume passage increases with height, while the downdraft stays at a similar strength at all heights. Figures 9c and d indicate that the model overestimated upward velocity at higher levels. The underestimation in the simulated horizontal wind speed at these levels shown by Figures 8 c and d could indicate that the modeled plume wasn't tilted downstream enough (was too vertical) so that the vertical wind component was overestimated while the horizontal one was underestimated. However a more vertical plume would result in delayed plume arrival at higher elevations. Figures 9 c) and d) indicate that this is not the case; the timing of the model updraft velocity peaks is captured correctly at 28 and 43 m AGL. It is more likely that this discrepancy between measured and simulated vertical velocities at upper levels is that the model overestimated the fire front depth, the amount of total heat released into the atmosphere, and therefore the plume updraft speeds. At low elevations, for reasons just discussed, simulated updraft velocities are numerically limited, so they match well with observations. At higher elevations, the model responds more freely to the excessive heating by increasing the vertical velocity within the plume.

a) Vertical velocity 2m main tower

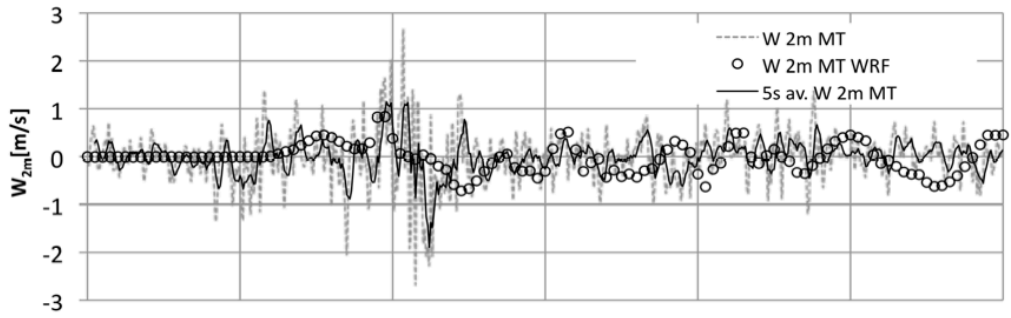

b) Vertical velocity 10m main tower

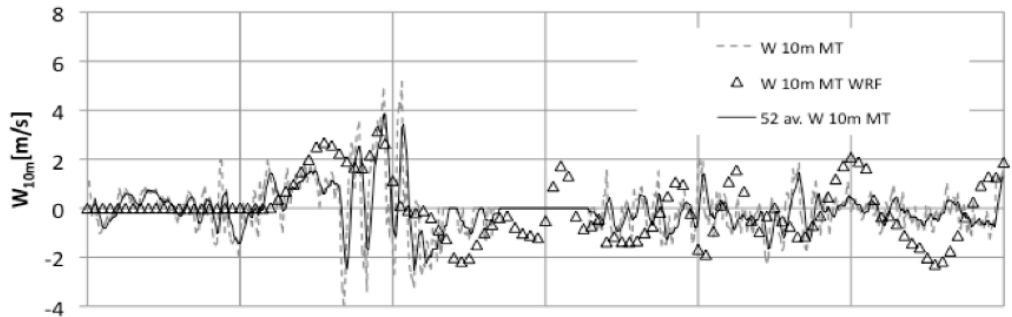

c) Vertical velocity 28m main tower

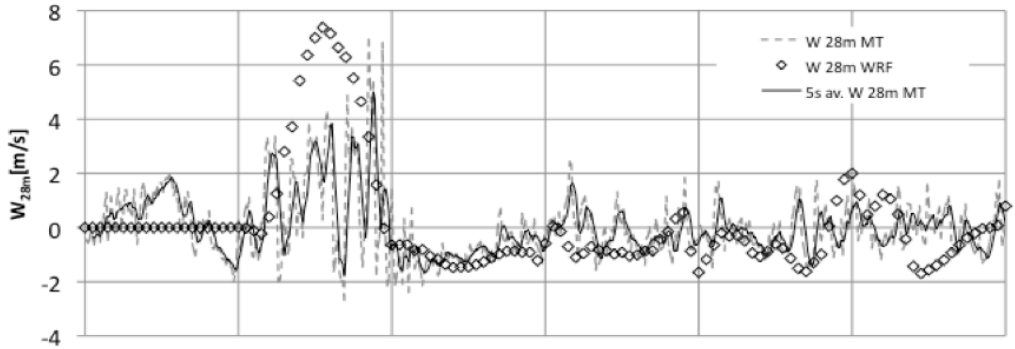

d) Vertical velocity 42m main tower

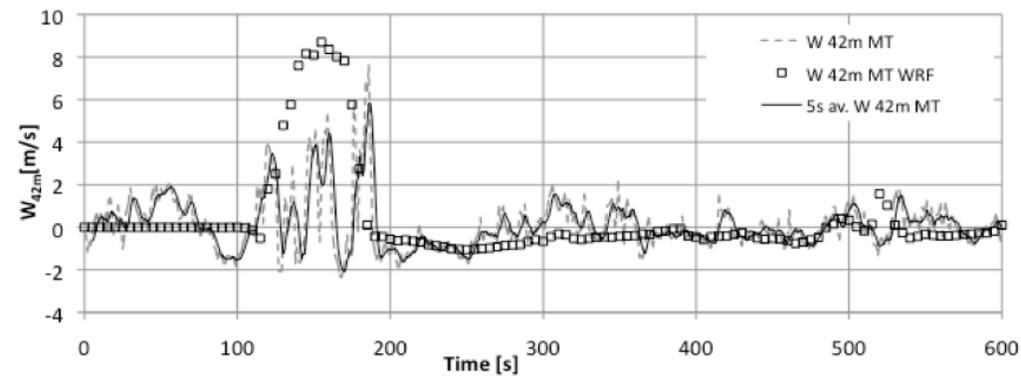

Figure 9. As in Figure 8 except for vertical wind speed (w).

*5.3.3. Spatial structure of the fire-induced flow.*

Based on the good agreement between FireFlux observations and WRF-Sfire results seen in Figures 4 to 9, a more detailed analysis of the possible dynamics responsible for FireFlux behavior as the fire passed the MT and ST may be attempted using the WRF-Sfire simulation. Here model flow properties $w$, the vertical $z$ velocity component, and $|\mathbf{V}_h|$, the magnitude of the horizontal wind velocity are examined, along with the following wind features:

$$\delta = \frac{\partial u}{\partial x} + \frac{\partial v}{\partial y},$$

the divergence in the horizontal *x-y* flow, and

$$\zeta^x = \frac{\partial w}{\partial y} - \frac{\partial v}{\partial z},$$

the *x* component of vorticity due to the development of shear in the *y-z* flow. Here *u,v,w* are the *x,y,z* components of the flow. The separation or coming together of flow parcels in the *x-y* plane is described by $\delta$, where $\delta > 0$ signifies divergence and $\delta < 0$ signifies convergence of flow parcels. The spin or rotation of flow parcels in the *y-z* plane is described by $\zeta^x$, where $\zeta^x > 0$ signifies cyclonic or counter-clockwise rotation and $\zeta^x < 0$ signifies anticyclonic or clockwise rotation of flow parcels. Figures 10 and 14 are *x-y* cross sections that illustrate WRF-Sfire behavior at 3 m AGL (the second height level in the model simulation) at two times: 3:45 (min:s) when the fire front reached the MT; and 7:45 (min:s) when the fire line reached the ST. Figures 11 and 15 are *y-z* cross sections that illustrate the WRF-Sfire behavior at *x*=465m at these two times.

Figure 10 shows all of the flow features described by Clements et al (2007) for 3:40. As the fire front approached the MT the surface wind speed more than tripled, and before the horizontal wind increase, there was a brief period of calm associated with horizontal convergence ahead of the fire line that coincided with increased vertical motion. Wind vectors in Figure 10 show clearly how, just ahead of the MT and the fire head, the direction and speed of the horizontal wind changed from ambient wind conditions of mainly northerly flow of approximately 3 m s$^{-1}$ to the almost reverse direction and almost calm wind conditions. The model wind behavior is very similar to the wind behavior seen in the Linn and Cunningham (2005) FIRETEC simulation of a 100 m long grass fire line in low (1 m s$^{-1}$) ambient wind conditions (their Figure 2). Figure 10 (b), (c) and (d) shows, respectively, considerable horizontal divergence, large horizontal wind speeds (10 to 12 m s$^{-1}$), and significant downdrafts just behind and along entire leading edge of the fire front. Horizontal convergence and vertical velocity are most significant immediately out ahead of the fire front. Convergence in the horizontal wind is strongest at the base of the narrow updraft. At the time of FFP, in agreement with observations, the WRF-Sfire horizontal wind speeds increased due to fire-induced circulations, while background winds outside the burn perimeter remained constant.

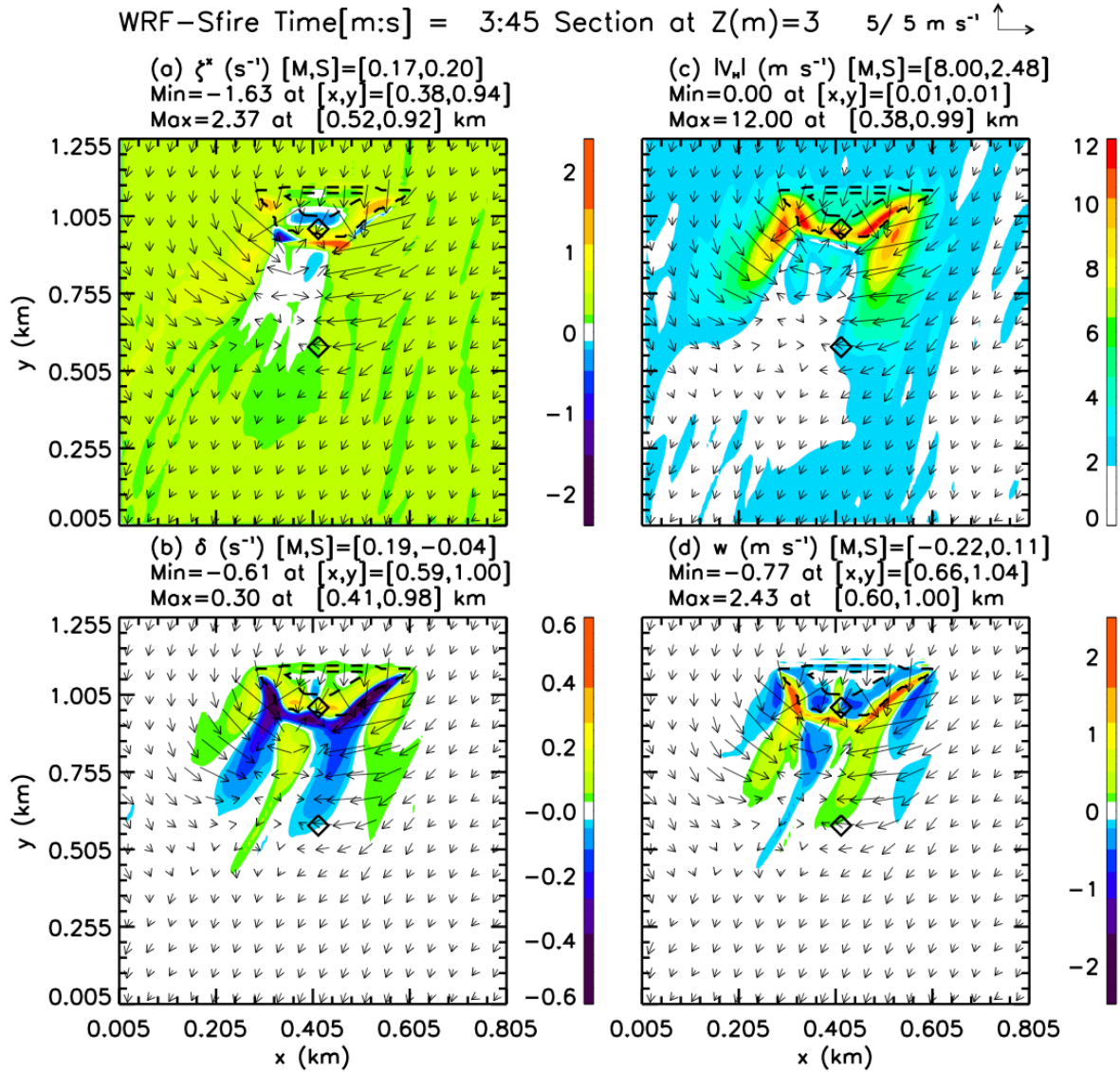

Figure 10. Horizontal cross sections for 3 m AGL of (a) horizontal $x$ vorticity $\zeta^x$ (s$^{-1}$), (b) horizontal divergence $\delta$ (s$^{-1}$), (c) speed of horizontal wind $|V_H|$ (m s$^{-1}$), and (d) vertical velocity $w$ (m s$^{-1}$) at 3:45 [min:s] into the WRF-Sfire simulation. Magnitudes of each contour are indicated by colors in bar plots on the right. For each field, minimum and maximum values, plus their (*x,y*) positions on cross section are given. Vectors denote wind components in *x-y* plane where magnitude is scaled as indicated in top right corner of plot. Black dotted contour lines delineate the surface fire perimeter. Note that the (aspect) ratio between the height of each plot to its width is not equal to one. Plots show features lengthened in the *y* direction compared to the *x* direction. The (*x,y*) locations of the MT and ST are indicated by a black triangle and diamond, respectively.

Figure 10 displays additional structure to the flow. Figure 10 (a) indicates positive $\zeta^x$ at the MT location and the leading edge of the fire front, and negative $\zeta^x$ behind. Downstream flow features are associated with horizontally-oriented convective rolls. Out ahead of the fire head are divergence, weak horizontal wind, and downward motion, between strong convergence, significant horizontal wind speeds, and upward motion. The convergence out ahead of the fire front on either side of the fire head may be responsible, in part, for Clements et al (2007)'s observation that the convergence zone was farther ahead of the fire front than previously thought. The model shows the fire head moving towards the south south-west as it reaches the MT.

Figure 11 shows *y-z* cross sections through the MT and fire head at time 3:40. By comparing Figures 10 (a) and (d) to Figures 11 (a) and (d), is it seen that the significant counter-clockwise (clockwise) $\zeta^x$ ahead of (behind) the leading edge of the fire head coincides with $\partial w / \partial y < 0$ ($\partial w / \partial y < 0$) as part of the model plume's updraft (relatively weaker trailing downdraft).

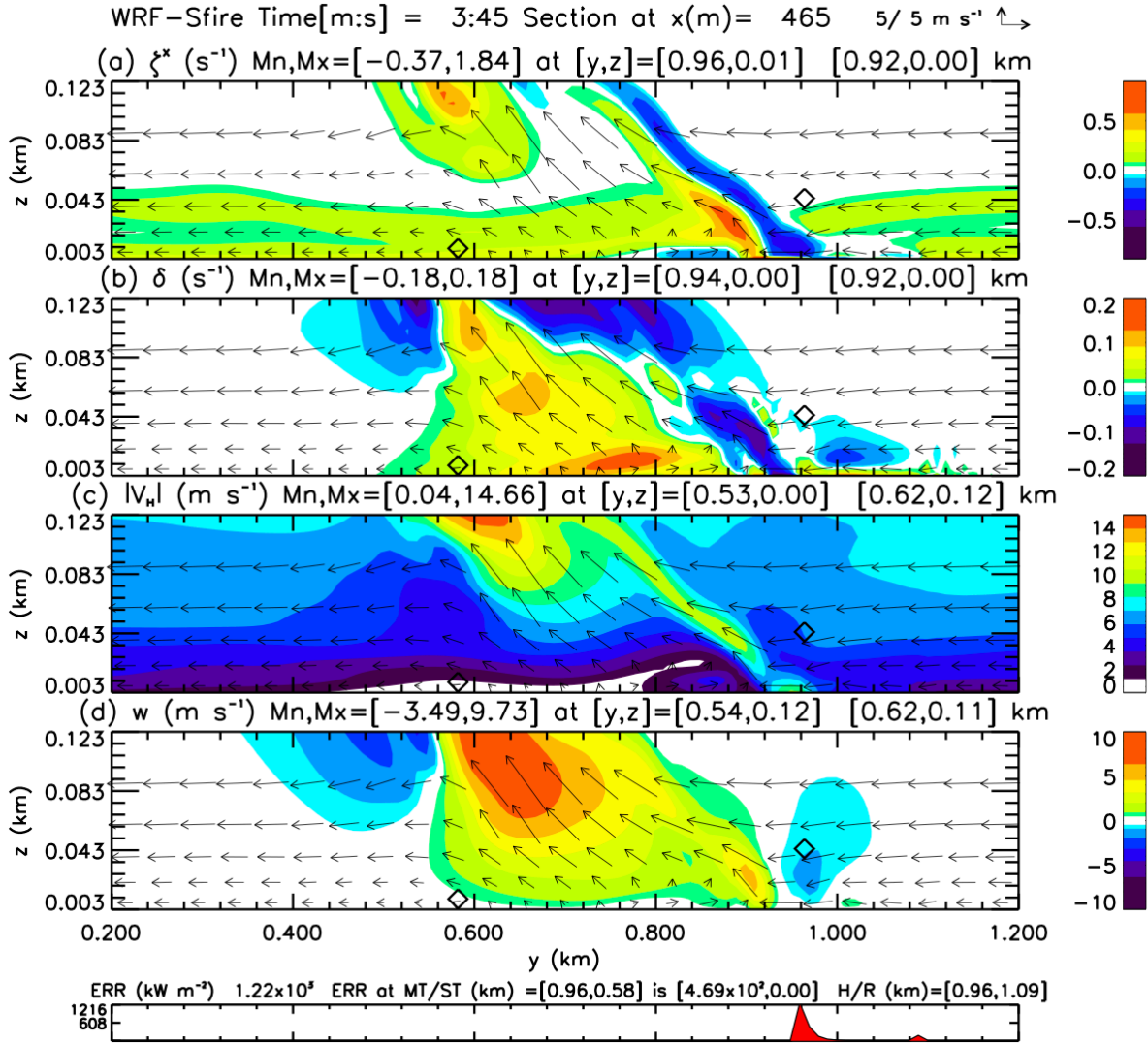

Figure 11. As in Figure 10 except for vertical *y-z* cross sections at *x* = 465 m. The bottom plot displays the Energy Release Rate (ERR) (kW m$^{-2}$) from the surface fire as a function of *y*. Maximum rear and head (R/H) distances (km) advanced by the fire are given, along with fire flux (ERR) values at the surface locations of the MT and ST. The top locations of the MT and ST are indicated by a black triangle and diamond, respectively.

As in Figure 10, Figure 11 shows, near the surface, divergence, weak to calm horizontal wind speeds, and weak vertical motion out ahead of the fire head.  The position and distribution of energy release rate (ERR) in the fire's head and rear line are seen in the bottom plot in Figure 11.  The maximum ERR is 861 kW m$^{-2}$ at the fire's front.  The wind vectors show winds shifting to undisturbed steady northerly flow once the fire front has passed.   Observations and model results indicate that just as the fire front passed the MT a period of downward motion occurred.  It is not clear that the downdraft rear of the fire front seen in Figures 10 (d) and 11 (d) is the cause of fire-induced winds as suggested by Sun et al (2006) and discussed by Clements et al (2007); it may be subsidence developing in response to the fire plume's sudden and strong convective updraft.  Both observations (Clements et al 2007) and model results (Figure 11) report the largest wind speeds occurred in upper-most plume level that was measured by the MT.   In Figure 11 the strongest vertical motion, horizontal wind speeds, and convergence occur at approximately 0.11 km AGL.

Clements et al (2007) and Figure 11a suggest a horizontal vortex immediately in front of the fire front at the MT.  Clements et al (2007) also describe soot particles (seen in video and time-lapse photography) dropping out in front of the head fire during the fire passage at the MT. Figure 11a indicates two regions of counter-clockwise rotation: a weaker one at upper levels near 0.12 km AGL, and a stronger one at the surface just downstream of the fire front at $y$=0.96 km.  It may be that the soot particles observed by Clements et al (2007) were entrained into the plume by the stronger surface horizontal vortex, carried up into the plume by this circulation, and then dropped out downstream of the fire.

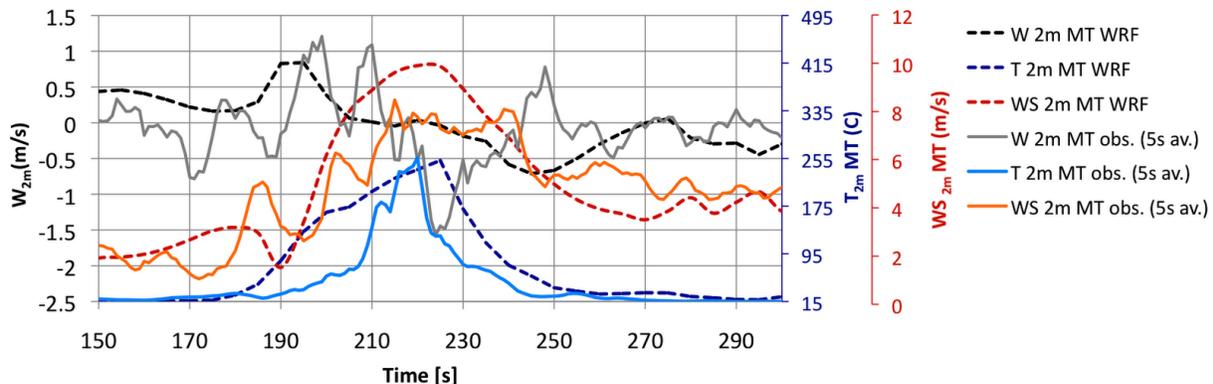

Figure 12. Time series from the MT of the simulated (dashed lines) and observed (solid lines) updraft velocity (W), temperature (T) and horizontal wind speed (WS) at 2m.  Observational results are presented as 5s averages of the original 1Hz data.

Close-ups of model results and observations of temperature and $w$ values during FFP at the MT are displayed in Figures 12 and 13.  Peaks in the observed and simulated vertical velocity (Figure 12; grey solid and dashed black lines, respectively) arrive earlier at the MT than peaks in observed and simulated temperature (Figure 12; solid and dashed blue lines, respectively).  Figure 13 shows that the WRF-Sfire updraft core is situated ahead of the fire front, whose position is identified by the maximum in the ERR.  The strongest surface convergence is associated with calm surface wind wind speed and located at the base of the plume's updraft, and both model results and observations suggest that these features are located ahead of, not in or above, the fire's head. Because of the downstream shift, ahead of the fire front, by convergence in the horizontal flow and associated upward motion, fire spread is driven by a local fire-induced wind (Figure 12; dashed red and solid orange lines) of much greater magnitude than the ambient one. Figure 12 shows that peaks in the simulated wind speed (dashed red line) and temperature (dashed blue line) are collocated. Strong surface winds cross the fire line, advecting fire-heated air downwind, where the warmed, buoyant air converges to form the base of the fire's plume. Note that the maximum ERR of ~2MW at the MT seen in Figure 13 is the WRF-Sfire instantaneous fire-grid mesh averaged value. Using 2m AGL thermocouple and vertical wind measurements, Clements et al (2007) estimated 1 MW m$^{-2}$ as a heat flux maximum. Note that the previous atmospheric grid-averaged ERR of ~ 1.216 MW compared to the 2MW fire-mesh ERR of 2MW indicates the sensitivity of the magnitude of model properties to grid-volume averaging.

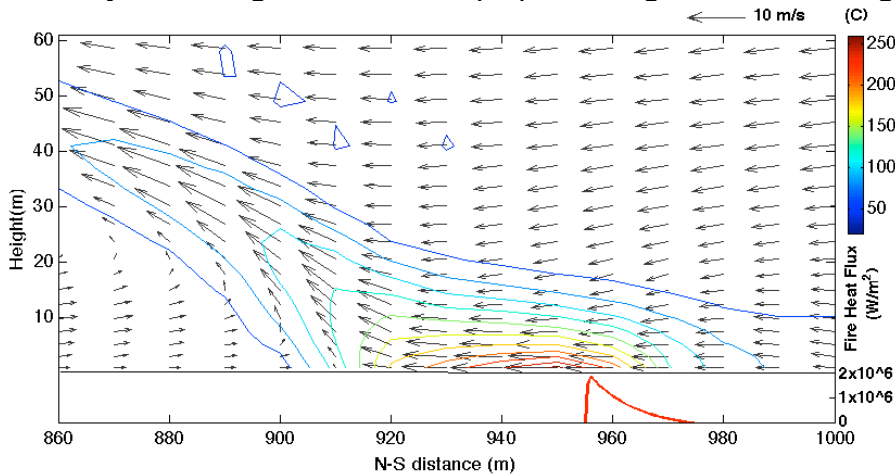

Figure 13. Vertical *y-z* cross section at *x*=465m, 225 s into simulation. Vectors denote wind components in *y-z* plane where magnitude is scaled as indicated in top right corner of plot. Contour lines represent air temperature (deg C), and the magnitude of each contour line is indicated by the colorbar on the right side of the plot. The red thick line shows the ERR (W m$^{-2}$) computed on the fire grid.

Figure 14 indicates that the WRF-Sfire fire head continues to move towards the south-west, and the model fire reaches the ST at 7:45 [min:s]. Figure 14 (b), (c) and (d) show, respectively, considerable horizontal divergence, large horizontal wind speeds (up to 19 m s$^{-1}$) and updrafts along and ahead of the leading edge of the model fire front. Convergence in the horizontal wind is strongest at the base of two updrafts positioned immediately out ahead of the fire front. The simulation shows the increased depth of the fire front and the fire, along with the winds in the south-east portion of the fire domain veering to the south-west. As the model fire front approaches the ST, the fire-induced flow develops flow features not seen at the MT at 3:45 (Figure 10). Wind vectors show clearly how, out ahead of the ST and the fire front, the horizontal wind is extremely turbulent and changed considerably from ambient wind conditions. This model wind behavior is very similar to the wind behavior seen in the Linn and Cunningham (2005) FIRETEC simulation of a 100 m long grass fire line in similar ambient (3 m s$^{-1}$) wind conditions (their Figure 3). Figure 14 shows complex patterns to $\zeta^x$, $\delta$, and *w*, not just out ahead of the fire, but over the entire area enclosed by the fire perimeter. There are alternating strips or streaks of up/down vertical motion coincident with convergence/divergence in the horizontal flow field. These appear to be organized horizontal rolls or eddies embedded in the burning area and aligned with the mainly northerly background flow, similar to the convective instabilities known as "cloud streets" that are common in the atmosphere (Brown 1980; Etling and Brown 1993). It should be noted that these fire "streets" did not develop until the Rothermel default no-wind fire ROS was increased from 0.02 m s$^{-1}$ to 0.1 m s$^{-1}$. There are no FireFlux data to validate this result, but this flow pattern is similar to the convective and radiative heating patterns seen in the Cunningham and Linn (2007) FIRETEC simulations of 100 m long grass fire lines (their Figure 4). These model results suggest that the heat released by actively moving fire flanks and back is essential to the production of these dynamic "fingers."

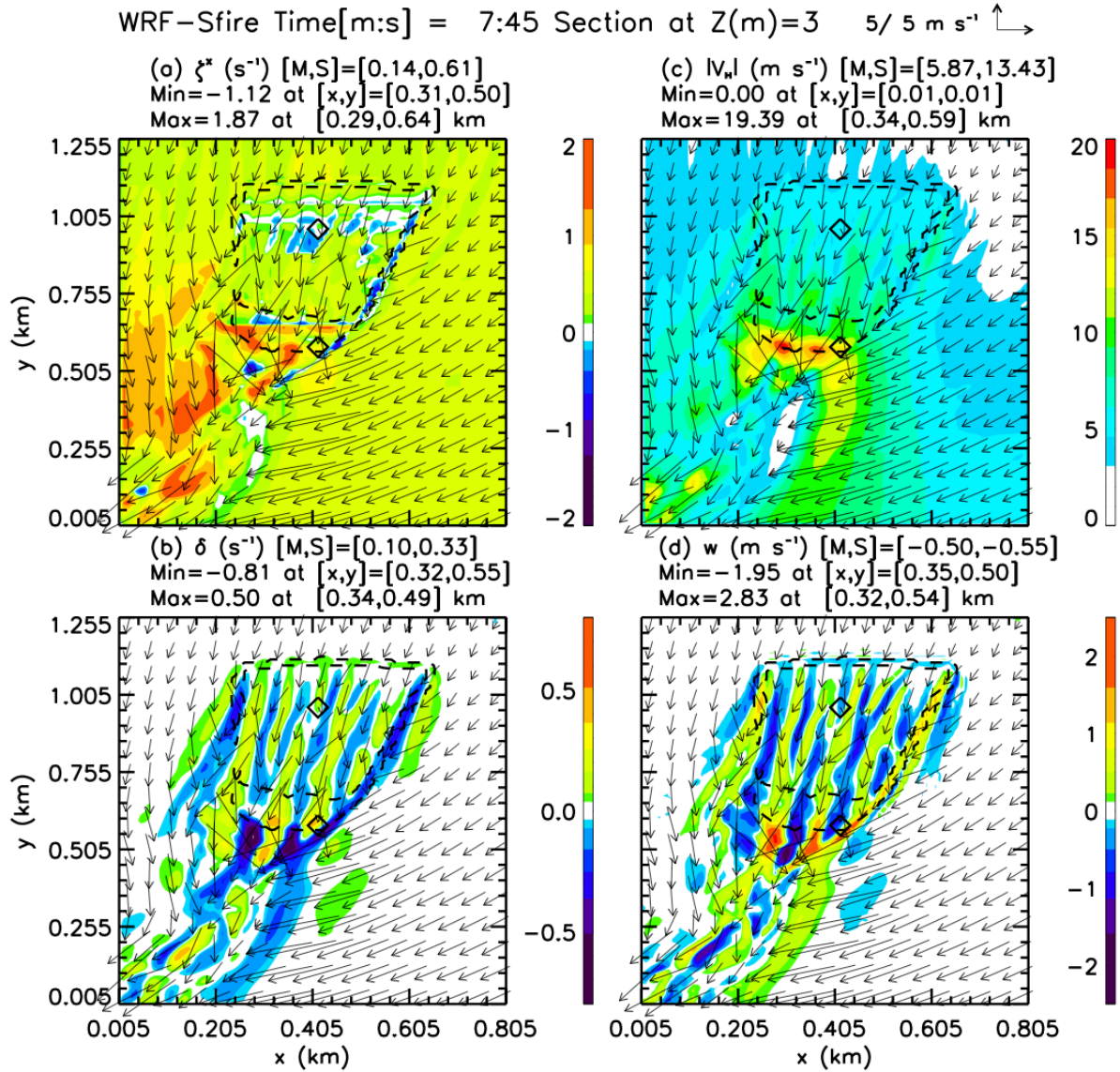

Figure 14. As in Figure 10 except for 7:45 [min:s] into the WRF-Sfire simulation.

Figure 15 shows *y-z* cross sections through the ST and fire head at time 7:45. As before, significant counter-clockwise (clockwise) $\zeta^x$ ahead of (behind) the leading edge of the fire head coincides with $\partial w/\partial y > 0$ ($\partial w/\partial y > 0$) as part of the model plume's updraft (relatively weaker trailing downdraft). The wind vectors do not show winds shifting to undisturbed steady northerly flow once the fire front has passed. Between the front and back fire lines, at 0.58 and 1.12 km in the *x* direction, respectively, flow is disturbed in the region of the fire showing what is likely the result of the convective instabilities or "fingering" seen in Figure 14. The model results indicate that, just as the fire front passed the ST, a period of downward motion occurred. The position and distribution of heating rates in the fire's head and rear line are seen in the bottom plot in the figure. Averaged on the WRF atmospheric grid mesh, the maximum ERR (Energy Release Rate) was 1045 kW m$^{-2}$ at the fire's front (bottom plot in Figure 15).

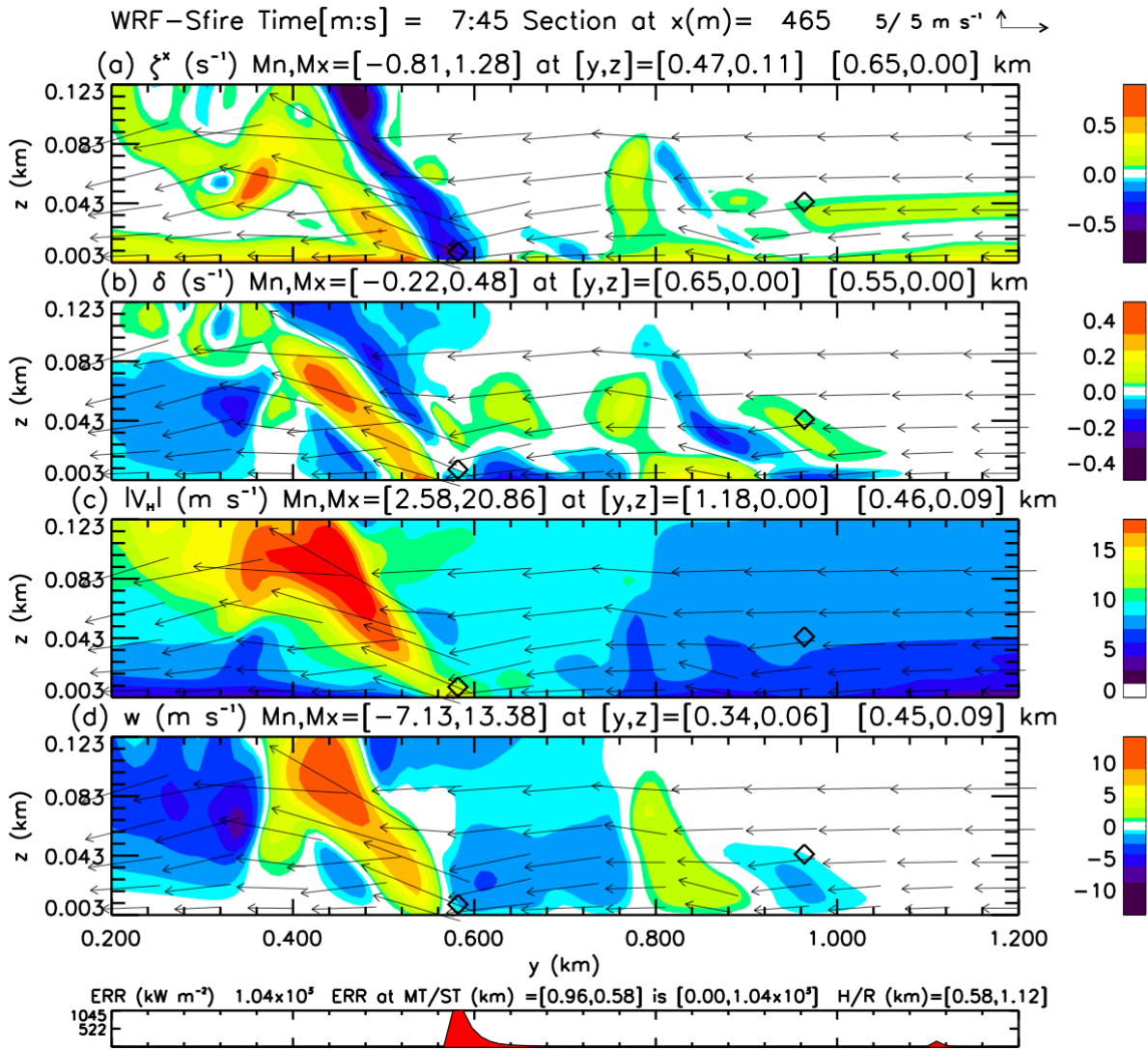

Figure 15. As in Figure 11 except for 7:45 [min:s] into the WRF-Sfire simulation.

As before at 3:45, wind speeds are largest at upper levels in the plume. Figure 15 shows the strongest vertical motion and horizontal wind at approximately 0.45 and .46 km AGL. Although there are no FireFlux data to validate these ST model results, they are consistent with the plume and fire behavior seen in Figure 11 for the MT. Model results (not shown) indicate that maximum vertical wind speeds are always found below 400 m AGL, while the largest vertical extent of the plume is approximately 800 m AGL.

## 6. Discussion

The results in Section 5 indicate that overall the agreement between WRF-Sfire and FireFlux was relatively good. It appears the WRF-Sfire simulated well the evolution of primary flow features in the FireFlux plume. In Section 4, it is seen that a few adjustments to WRF-Sfire were necessary to match FireFlux behavior, especially in the early phase of the fire. Here the importance of these adjustments to WRF-Sfire as a predictor of wildfire behavior is discussed, followed by suggestions for the design of future field campaigns that are required to develop and validate numerical coupled atmosphere-fire prediction models such as WRF-Sfire.

It is understood that, after initial ignition, wildfires experience an "acceleration" or growth phase, before reaching an "equilibrium" or quasi-steady rate-of-spread (Cheney and Gould, 1997). WRF-Sfire was coded therefore to take this fire growth phase into account, using arrival time at the MT as a guide. By taking the fire's initial growth phase into account, the simulated fire propagation times to the MT compared very well to the observations.

Current operational fire-spread models are formulated for head-fire propagation where, typically, a single generalized default no-wind spread-rate is applied along both the fire's back and flanks. But as Mell et al (2007) demonstrates, there is no general flank- or back- fire spread rate; modeling the evolution of the entire fire line is a greater challenge, due to the different spread mechanisms, than modeling the behavior of just the head fire. Rothermel's default no-wind rate spread value for the grass of properties shown in Table 1 is 0.02 m s$^{-1}$, which ensures essentially zero spread along the back or flanks of a fire. Used in preliminary WRF-Sfire runs, this no-wind value did not provide good agreement with the FireFlux fire line's arrival at the ST. The fire front was so skewed that the ST was passed by a fire flank rather than its head. Therefore, in order to achieve realism of the FFP, this value was increased to 0.1 m s$^{-1}$. However, this important parameter impacts the heat release rate, and the result in this study was active flank- and back-fire spread with discernible consequences for fire plume properties and behavior. If flanking fire and backing fire spread are due to different mechanisms, then it is in general not appropriate to apply a single no-wind fire-spread value as done in the Rothermel fire-spread formulation. It was not possible however to determine, using available FireFlux observations, if the simulated flank and back-fire spread rates reproduced accurately the entire fire perimeter spread or not. It may be worthwhile to investigate the use of fire-spread formulations other than Rothermel's in WRF-Sfire, such as Balbi et al (2007), that require a relatively small number of input parameters and provide a variable no-wind fire spread rate depending on these parameters.

A second fire model parameter that impacts heat release is the fuel depth. Clements et al (2008) estimated 1.5 m as the depth of the grass fuel, whereas in this study, in order to produce agreement between simulated and observed fire behavior, a fuel depth of 1.35 m was used. The Rothermel fire-spread model is particularly sensitive to fuel properties such as moisture content (Jolly 2007) and the fuel depth. Again, this result suggests that fire growth models other than Rothermel's should be tested in WRF-Sfire.

A third important fire model feature is the e-folding extinction depth used to parametrize the absorption of sensible, latent, and radiant heat from the fire's combustion into the surface layers of WRF. In this study the flame length estimate of 5.1 m by Clements et al (2007) was used to set the extinction depth to 6 m, with the result being that the WRF-Sfire vertical profile of temperature taken at the main tower was in good agreement with FireFlux observations, whereas the vertical profile of vertical velocity shows WRF-Sfire values larger than those observed. This relatively good temperature agreement suggests that efforts to explicitly distinguish between or model the different modes of fire-atmosphere heat transfer (conductive, convective, radiative) may not, at substantially greater computational cost, provide substantially better plume temperature prediction.

This study provides the opportunity to suggest the design of future field campaigns used to evaluate or validate numerical wildfire models such as WRF-Sfire. In addition to the observing procedures to measure winds, temperature, humidity, and surface pressure, described in Clements et al (2007, 2008) and Clements (2010), the following are suggestions for field campaign protocol based on the results of this study.

The experimental layout needs to be measured carefully for spatial dimensions, any special geographic features, and tower and equipment positioning. This suggestion is based on the finding that the evaluation of simulated fire was sensitive to the accuracy of these features and their locations in the WRF-Sfire model domain. Positioning done with GPS ranges in accuracy from 10-30 cm to (more typical) 1-5 m, depending on the GPS receiver.

The position of the initial fire line should be clearly marked and reported, and the timing of the walking-ignition well determined. In addition, to ensure uncomplicated initial fire line behavior, the initial fire line should be as perpendicular as possible to, ideally, a directionally-steady background wind. These suggestions are based on the observation that the evolution of simulated fire appears to be sensitive to any asymmetry in the timing and positioning of the walking-ignition and prevailing winds.

The rate of spread, flame length, and heat release per unit area were estimated in FireFlux (Clements et al, 2007) using the Behave-Plus application (Andrews et al 2005) and the weather observations at the time of the burn. In addition therefore, before a burn, it is recommended that the WRF system and the WRF-Sfire be run separately in the LES (Large Eddy Simulation) mode to provide, respectively, initial fine-scale atmospheric no-fire and fire data for the area of a field experiment to help with micro-siting and utilization of instrumentation (e.g., number and location of measurement towers, measurement levels, measurement frequency, etc). Before a burn, ideally, efforts should be made to gather in-situ high-frequency fine-scale measurements of momentum fluxes, turbulence, and wind that are needed to verify the no-fire wind features predicted by WRF-LES in the ABL. WRF-LES wind forecasting and nowcasting abilities would be evaluated with comparisons between ensemble averages of the LES turbulent flow results and these field measurements. Note that in this study observations at greater than 1-Hz sampling rate were not needed or used to evaluate WRF-Sfire.

A LES is inherently unsteady. There are studies, for example Chow and Street (2009), that suggest that, for a LES simulation to predict successfully both mean flow and turbulence in the ABL, it should be provided with inflow conditions based on a separate, predetermined turbulent flow database. The ensemble averages of the no-fire WRF-LES and field data turbulent flow results would be used for this purpose.

The placement of the observing platforms relative to the initial fire line and wind field is important. The tower arrangement in FireFlux was intended to capture the flow and temperature fields at the fire-atmosphere interface as the fire front traveled with the wind and passed each tower consecutively (Clements et al 2008). It is recommended that taller (main) instrumented towers be placed farther downstream from smaller (shorter) towers. This layout is different from the one used in FireFlux and is based on the observation that the fire line's behavior and plume are, respectively, relatively simple and small in the early stages, growing more complex and taller with time. Clements et al (2007) notes that an array of towers aligned east-west would have provided a better description of the surface flow and verification by direct observation of the fire-induced flow features associated with the combustion-zone winds.

Although a tethersonde system in tower mode with five sondes was deployed during FireFlux, data during the fire are missing due to the loss of the tethered balloon as a result of strong vertical downdrafts during the initial plume impingement on the balloon. These data provide the above-tower (i.e., upper-level) vertical structure of temperature, humidity, and wind in the fire plume, and are especially valuable for a model validation study. Based on WRF-Sfire results, the maximum plume height was estimated at 800 m AGL, an ABL depth that only a tethersonde system can measure. It is known now from the FireFlux experience just how strong the tether for the tethersonde system needs to be.

A radiosonde launched on site just before the burn, instead of a few hours earlier, would be most useful for documenting the background atmospheric conditions. Even without any large-scale synoptic forcing, both wind and temperature can change in just a few hours as part of the normal diurnal cycle or topography-influenced meteorology. Basic, portable, weather stations located upwind and outside the burn perimeter would also provide background meteorological measurements before, up to, and during the burn.

Multiple digital infrared video and visible SLR cameras can be employed to document smoke and flames. Using a still exploratory method, Clark et al (1999) show how it is possible to calculate convective-scale velocities and heat fluxes from infrared imagery. Doppler lidar (Banta et al. 1992) can also be used to observe the finer-scale kinematics of fire plumes.

The spread of the entire fire perimeter should be measured accurately. In FireFlux, even though orange markers were placed in the fuel at 10-m intervals from 50 m north to 300 m south of the main tower to aid in head-fire spread rate determination, this information was not enough to evaluate the size and shape of the entire fire perimeter as the fire evolved. Aerial video (shot from a helicopter) and time-lapse photography can provide information on perimeter spread, but ideally this information should be supplemented with measurements from a surface-based thermocouple array. In FireFlux, soil temperature thermocouples were buried 3 and 10 cm below the surface, but these were placed only near the base of the MT (Clements et al 2008). Thermocouples capable of measuring temperatures up to 1200 Celsius, and housed in a (plastic) unit, buried just below (5 cm or so for grass fires, 10 cm for higher intensity burns) the surface, can be used to determine fire spread. Thermocouples with higher temperature insulation last for multiple fires. Also useful would be thermocouples set to trigger at some earlier point in the day to monitor ambient temperatures at 1 HZ for as long as 12 hours.

The FireFlux burn lasted for approximately 17 minutes. As described in Cheney and Gould (1997), and references therein, the typical fire growth curve for a fire burning under fairly stable fuel moisture and wind conditions takes approximately 30 minutes before reaching a quasi-steady rate-of-spread. Ideally, measurements from burns lasting at least that long would be very valuable for evaluating numerical fire behavior prediction models such as WRF-Sfire.

## 7. Concluding Remarks

In this study, FireFlux observations (Clements et al 2007, 2008; Clements 2010) --- the first comprehensive set of in situ measurements of turbulence and dynamics in an experimental wildland grassfire --- were used to evaluate and improve the forecast capabilities of WRF-Sfire, and the different model changes made to WRF-Sfire have been described. Missing observations in FireFlux made many direct model/observations comparisons difficult, and a more complete evaluation of the WRF-Sfire's performance for predicted properties of surface pressure, evolution of wind fields, plume properties, and surface fire perimeter spread is required. Based on the comparisons that were possible, the overall agreement between model results and tower measurements in terms of matching head-fire spread rates, vertical profiles of temperature, vertical wind and horizontal wind speeds, is encouraging. A more intensive observational field campaign should be conducted, and based on the FireFlux experience and the results of this study, suggestions are made for optimal experimental pre-planning, design, and execution of such a campaign.

A long-term goal is to develop and test WRF-Sfire for operational real-time wildfire prediction. Meanwhile, the level of agreement between WRF-Sfire simulated results and FireFlux observations suggests that it would be feasible to test and use WRF-Sfire for wildfire management in prescribed burns, smoke dispersion, or emergency evacuation, under wind and terrain conditions similar to FireFlux.

ACKNOWLEDGMENTS. This research was supported in part by Department of Commerce, National Institute of Standards and Technology (NIST), Fire Research Grants Program, Grant 60NANB7D6144. The University of Houston Houston Coastal Center is acknowledged for financial support for the FireFlux experiment. A gratis grant of computer time from the Center for High Performance Computing, University of Utah, is gratefully acknowledged.

**Tables:**

Table 1. Details of the numerical setup used for the FireFlux simulation.

| Simulation type | LES (Large Eddy Simulation) |
|---|---|
| Horizontal domain size | 1000 m x 1600 m |
| Atmospheric mesh | 160x100x80 |
| Horizontal resolution (atmospheric mesh) | 10 m |
| Model top | 1200 m |
| Vertical resolution (atmospheric mesh) | From 2 m at the surface to 33.75 m at the model top |
| Fire mesh | 3200x2000 |
| Horizontal resolution (fire mesh) | 0.5 m |
| Simulation length | 20 min |
| Time step | 0.02 s |
| Sub-grid scale closure | 1.5 TKE |
| Lateral boundary conditions | Open |
| Surface layer physics | Monin-Obukhov similarity theory (sf_sfclay_phys=1) |
| Land Surface Model | SLAB 5-layer MM5 model (sf_surface_physics=1) |
| Ignition time | 12:43:30 CST |
| Length of the western ignition line | 170m |
| Duration of the western ignition | 153s |
| Length of the eastern ignition line | 215m |
| Duration the eastern ignition line | 163s |
| Thickness of the ignition line | 1m |
| Heat extinction depth | 6m |
| Default (no wind, no slope) rate of spread | 0.1 m/s |
| Fuel depth | 1.35m |
| Ground fuel moisture | 18% |
| Fuel load | 1.08 kg/m$^2$ |
| Fuel type of the burnt area | 3 (Tall grass) |